\documentclass[referee,a4paper,12pt,traditabstract]{swsc} 

\usepackage{graphicx}
\usepackage{txfonts}
\usepackage{subfigure}
\usepackage{epstopdf}
\usepackage[displaymath,mathlines]{lineno}
\usepackage[authoryear,round]{natbib}
\usepackage[backref]{hyperref}
\usepackage{url}

\hypersetup{colorlinks=true,citecolor=cyan,urlcolor=cyan,linkcolor=blue}

\begin{document}

   \title{Solar flare forecasting using morphological properties of sunspot groups}

   \titlerunning{Solar flare forecasting}

   \authorrunning{Falco, Costa and Romano}

   \author{Mariachiara Falco\inst{1},
					 Pierfrancesco Costa\inst{1}
          \and
          Paolo Romano\inst{1}
          }

   \institute{INAF - Osservatorio Astrofisico di Catania, Via S. Sofia 78, 95123, Catania (Italy)\\
              \email{\href{mailto:mariachiara.falco@inaf.it}{mariachiara.falco@inaf.it}}
         }


   \abstract 
   {We describe a new tool developed for solar flare forecasting on the base of some sunspot group properties. Assuming that the flare frequency follows the Poisson statistics, this tool uses a database containing the morphological characteristics of the suspot groups daily observed by the Equatorial Spar of INAF $-$ Catania Astrophysical Observatory since January 2002 up today. By means of a linear combination of the flare rates computed on the base of some properties of the sunspot groups, like area, number of pores and sunspots, Zurich class, relative importance between leading spot and density of the sunspot population, and type of penumbra of the main sunspot, we determine the probability percentages that a flare of a particular energy range may occur. Comparing our forecasts with the flares registered by GOES satellites in the 1$-$8 \AA{} X$-$ray band during the subsequent 24 hrs we measured the performance of our method. We found that this method, which combines some morphological parameters and a statistical technique, has the best performances for the strongest events, which are more interesting for their implications in the Earth environment.}


   \maketitle

\section{Introduction}

The solar eruptions are the main manifestation of the magnetic activity of the Sun in the heliosphere with also an impact on Earth environment. The effects of these eruptions on the Earth and anthropological activities involve more and more economical interests, considering that the technology pervades many aspects of our life. Therefore, many efforts of the scientific community have been addressed to forecast solar eruptions with enough advance to prevent damages to technological systems.

Solar eruptions involve the whole solar atmosphere and are usually accompanied by sudden and intense variation in brightness, which are named flares. They are the manifestation of the release of free magnetic energy (see \citet{Benz17} for more details). Although their origin is usually located in the higher layers of the solar atmosphere, the magnetic field configuration suitable for the occurrence of these events is mostly driven by the photospheric evolution of the active regions (ARs). The emergence of new magnetic flux from the convection zone into the solar atmosphere and the rearrangement of the coronal field due to the horizontal photospheric displacements of the field line footpoints are two fundamental mechanisms which determine the AR configuration suitable for the flare occurrence (e.g., \citet{Romano07}, \citet{Romano15} and \citet{Romano18}). In particular, these mechanisms have greater effects when they occur near the magnetic polarity inversion line (PIL).

Many approaches to flare forecasting have been produced in last decades (see \citet{Barnes16} for an overview, see \citet{Kim19}, \citet{Korsos19}, \citet{Florios18} and \citet{Kontogiannis18} for more recent results). Although the configuration of the magnetic field in corona is fundamental for the trigger of flares, due to the difficulties to measure directly the magnetic field in the upper layers of the solar atmosphere, many of the flare forecasting methods are based on photospheric observations of the ARs.  

Using three photospheric magnetic parameters, i.e., the total unsigned magnetic flux, the length of the strong-gradient magnetic polarity inversion line, and the total magnetic energy dissipation, \citet{Yuan10} developed a method which obtained some of the best prediction results for the X-GOES class flares. This result confirms that the photospheric configuration of an AR provides already a good indication for the occurrence of the extreme events, which are the most interesting for their impact to our life. Another method to forecast strong flares considers the free magnetic energy obtained from the gradient of the neutral lines and the total unsigned flux measured at photospheric level \citep{Falconer08}. In fact, it has been confirmed that the magnetic flux close to the PIL is a good proxy for several methods (see \citet{Leka_Bar_03} and \citet{Schrijver07}). More recently, another forecasting method was proposed by \citet{Korsos15}. According to this method, the pre-flare behaviour of spot groups is investigated by the introduction of the weighted horizontal magnetic gradient ($WG_{M}$). This proxy enables the potential to forecast flares on the base of the magnetic gradient among all spots within an appropriately defined region close to the PIL (see also \citet{Korsos16}). The $WG_{M}$ allows to forecast the flare onset time and whether a flare is followed by another event within about 18 hr, with a good reliability.

The photospheric measures of the magnetic field can be also used  to produce a model of the coronal field and determine the connections between each pair of magnetic sources \citep{Barnes05}. A successful prediction method based on the coronal magnetic connectivity has been developed by \citet{Georgoulis07}. The computation of a parameter named {\em effective connected magnetic field strength} which is derived from a connectivity matrix describes the magnetic flux distribution of an AR and provide a probability of flare occurrence. Also this method seems efficient for major solar flares. 

Beside of the above mentioned methods there are also statistical methods which obtained good results: for example a method based on the flaring history of the observed ARs developed by \citet{Wheatland04}. This method assumes that the events obey a pawer$-$low frequency$-$size distribution and occur on short timescales following a Poisson process with a constant mean rate. This method appeared to be more accurate at predicting X events, which are more important contributors to space weather, than M events. Another forecasting method, performed on statistics base, has been proposed by \citet{Bloomfield12}. It uses the McIntosh group classification of sunspot groups (SGs) observed in photosphere and assumes that flares are Poisson-distributed processes. This method showed that Poisson probabilities perform comparably to other more complex prediction methods. 

A completely different approach to flare forecasting belongs to the Machine Learning framework. In this case, the properties utilized for
prediction (named features) are automatically extracted from observations and they are ranked according to their correlation with the event occurrence. Recent examples of machine learning results have been obtained by \citet{Florios18}, \citet{Inceoglu18}, \citet{Kim19}, and \citet{Camporeale2019}.

Recently, a very useful and complete comparison among some of the above mentioned flare forecasting methods has been performed by \citet{Barnes16},  reaching the conclusion that it may be possible to obtain the best prediction by combining a method which characterizes an AR by one or more parameters and uses a statistical technique. For this reason, in this paper we report the results obtained by the development of a new method which is based mainly on the Zurich classification of the SGs observed at photospheric level and assuming the Poisson statistics for the flare occurrence. We provide an estimation of the capability to host flares of a specified energy range for an AR characterized by a particular configuration, size and fragmentation \citep{Contarino09} . The data used to reach this goal are described in section 2. The method is explained in Section 3. Section 4 reports the validation process used to estimate the reliability of our forecastings. Section 5 contains some conclusions.

\section{Data description}
   
We used data collected daily by the Equatorial Spar of INAF$-$Catania Astrophysical Observatory (INAF$-$OACt) from January 2002 up today. When the weather conditions permit, this telescope allows the observation of the photosphere in White Light (WL) by a Cooke refractor (150mm/2230 mm) which projects a 24.5 cm diameter image of the Sun with a spatial resolution of about 1-2 arcsec depending on the seeing conditions. Every day, with a cadence of 24 hours, each SG visible on the disk is cataloged registering some information about the heliographic latitude and longitude of its baricenter, the number of sunspots and pores in the group, $SS$, the projected area in tens of millionths of the solar hemisphere, $AA$, the type of penumbra of the main sunspot of the group, $t1$, the relative importance between the leading spot and density of the sunspot population, $t2$ (see Table 2 of \citet{Ternullo06} for the details about this parameter) and the group type according to Zurich classification, $t3$ (see \citet{Zirin1998}).

\begin{figure}[h]
\centering
\includegraphics[scale=0.465]{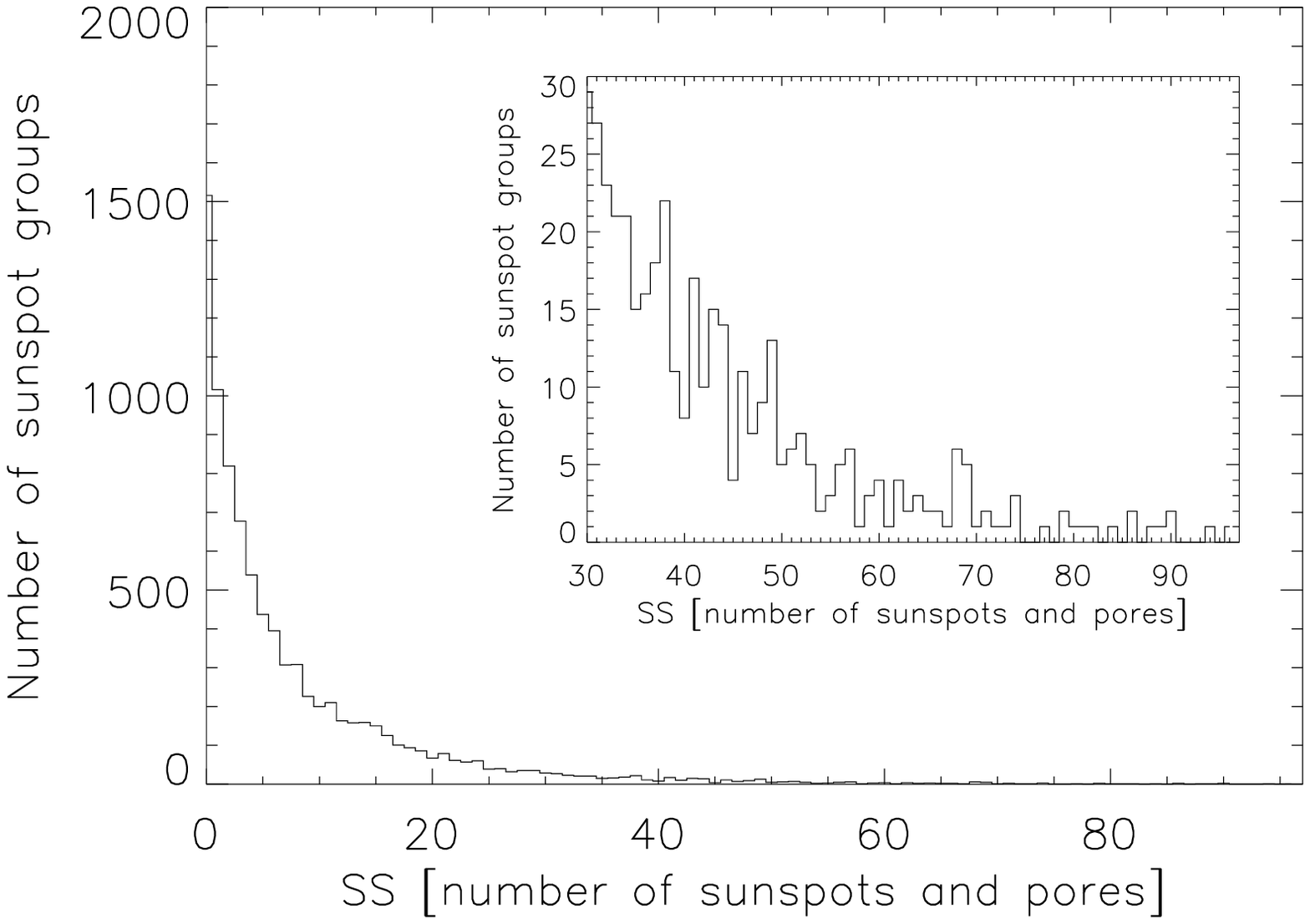}     
\includegraphics[scale=0.465]{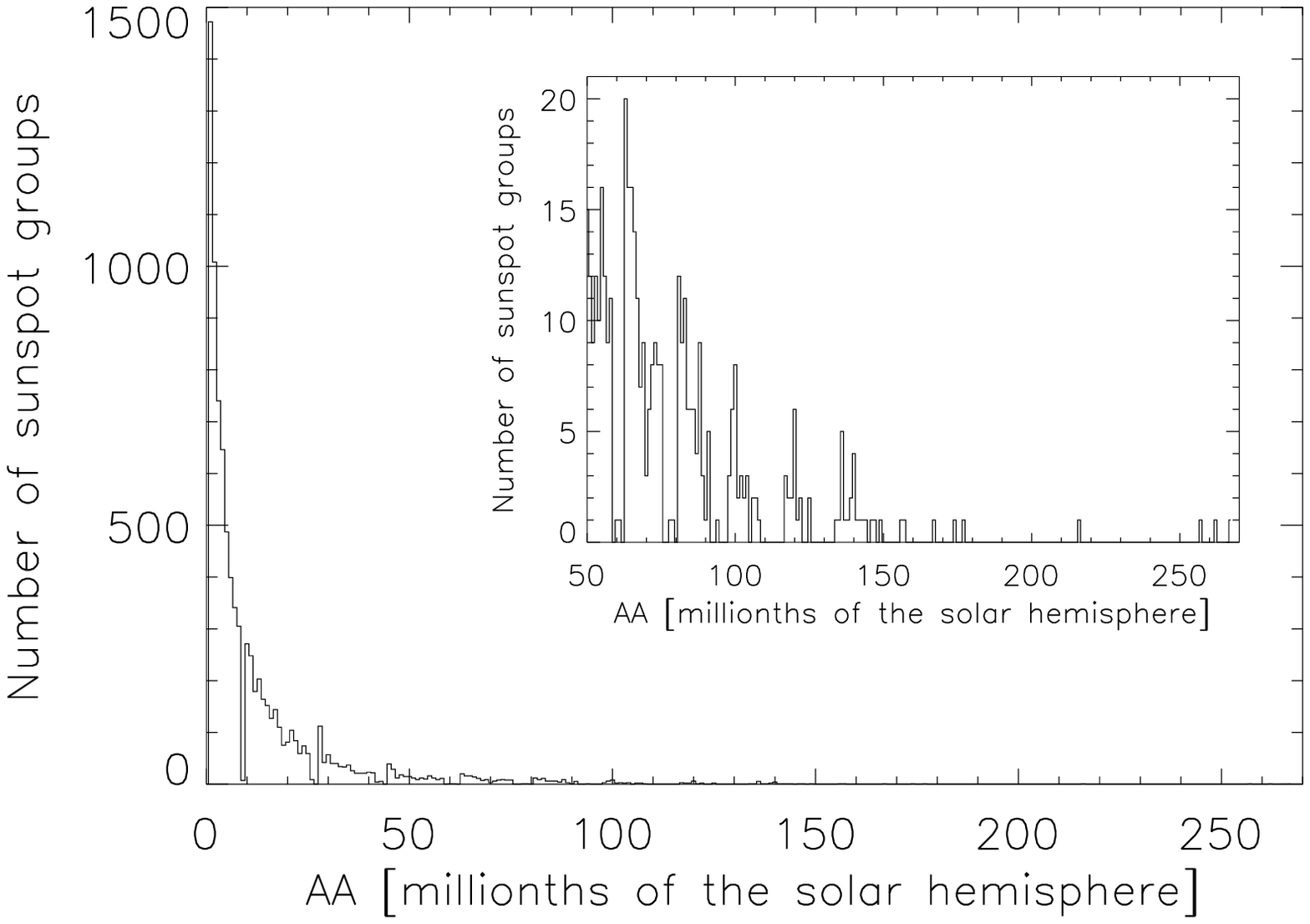}     
\caption{Distribution of the number of SGs as function of $SS$ (left panel, (a)) and $AA$ (right panel, (b)). In order to show the distribution for higher values of $SS$ and $AA$ we also reported the corresponding zoom of part of the plot in the top right space of each plot. \label{fig:res00}}
\end{figure} 

We considered 8598 SGs. While $t1$, $t2$, and $t3$ assume predefined values on the base of the classification of the SG morphology (see \citet{Ternullo06} for the details), $SS$ and $AA$ can assume a wide range of values on the base of the size and fragmentation of the groups. Their distribution in terms of the number of sunspots and pores is reported in the left panel of Figure \ref{fig:res00}. $SS$ ranges from 1 (single suspot or single pore) to 96 and we note a monotonic decrease of the number of SGs as the number of sunspots and pores increase. The distribution of the SGs on the base of their projected area is shown in the right panel of Figure \ref{fig:res00}. We can see that also the number of SGs is inversely proportional to their area, although we do not observe the same smoothed behavior of $SS$ distribution (compare the zoom plots reported inside the main plots of the left and right panels of Figure \ref{fig:res00}). The largest group observed during the considered time interval had an extension of 267 tens of millionths of the solar hemisphere. 

Table \ref{table0} contains the number of SGs per each morphological characteristic. We note that in our sample more than 43\% of the SGs (3711) were characterized by $t1$=2, i.e. by a symmetric main sunspot with a diameter less than 2.5$^{o}$. About 1500 SGs were formed by only pores ($t1$=0), while the number of groups characterized by other types of penumbra of the main sunspot were of the same order of magnitude. For the parameter $t2$ three are the main configurations: $t2$=0, corresponding to SGs formed only by pores or single sunspots, $t2$=1, corresponding to SGs formed by a leading largest sunspot and following smaller pores, and $t2$=4, corresponding to SGs formed by a leading largest sunspot and following smaller sunspots. $t3$ is based on the Zurich classification and shows that SGs of $C$ and $D$ type are the most diffused. Instead, SGs of $F$ and $G$ type, corresponding to the more extended and complex SGs are less diffused, although they are usually the preferential sites for the more energetic flares.

\begin{table}
\caption{Number of SGs per each morphological characteristic, t1,t2 and t3.}
\label{table0}
\begin{tabular}{|c|c|c|c|c|c|}     
\hline                   
$t_{1}$           & Number of      & $t_{2}$     & Number of        & $t_{3}$    &   Number of           \\
             & sunspot groups &        & sunspot groups   &       & sunspot groups        \\
\hline
0            &  1543          & 0      &    2511          &   A   &     652               \\

1            &   864          & 1      &    2280          &   B   &     898               \\

2            &  3711          & 2      &     473          &   C   &    1853               \\

3            &   957          & 3      &     301          &   D   &    2066               \\

4            &   776          & 4      &    2022          &   E  &     972               \\
                                            
5            &   743          & 5      &     866          &   F   &     289               \\
       
             &                & 6      &      51          &   G   &      17               \\
                                            
             &                & 7      &      44          &   H   &     327               \\
        
             &                & 8      &      44          &   J   &    1524               \\
         
             &                &  9     &       6          &       &                       \\
\hline                                             
        
\end{tabular}

\end{table}

We also used the flare records obtained by the GOES satellites, from January 2002 up today, and collected in the Space Weather Prediction Center Reports (https://www.swpc.noaa.gov/products/solar-and-geophysical-event-reports) to evaluate the performance of our forecasting method. In particular, we considered start time and class for each flare of C1.0 and greater observed by GOES. We neglected B-class flares.

\section{INAF$-$OACt forecasting method}

Our method to investigate the SG capability of hosting flares during the AR evolution assumes that the flare event frequency follows the Poisson statistics, as in \citet{Bloomfield12}. We compute the flare rates, $FRs$, of SGs characterized by specific values of the above mentioned five parameters: $SS$, $AA$, $t1$, $t2$ and $t3$.  While $t1$, $t2$, and $t3$ assume predefined values corresponding to different classes (according to the Zurich classification), $SS$ and $AA$ cover a wide range of values. Therefore, taking into account the distributions of the observed SGs as function of $SS$ and $AA$ (right and left panels of Figure \ref{fig:res00}), we divided the whole dataset into statistically significant and homogenous subsets (classes), considering 23 and 31 subsets for $SS$ and $AA$ values, respectively.

\begin{figure}[h]
\centering
\includegraphics[scale=0.20]{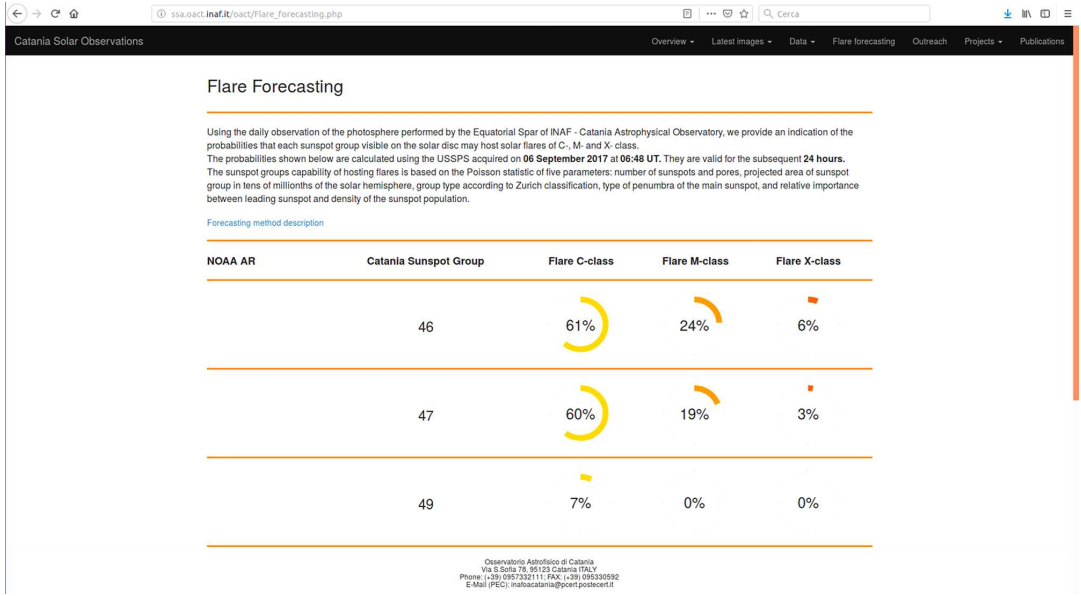}     
\caption{Screenshot of the INAF$-$OACt web page reporting the flare forecasting for September 6, 2017. \label{fig:res0}}
\end{figure}

For each specific value, $x$, of each generic parameter, $k$, we compute the $FR$, by the ratio between the number of SGs which hosted at least a flare, $Nf(x_{k})$, and the total number of SGs characterized by that value of the parameter $N(x_{k})$:
\begin{equation}
\centering
FR_{k}(x_{k})=\frac{{N_{f}(x_{k})}}{N(x_{k})}.
\label{eq1}
\end{equation}

Then we compute the average among the flare rates for all parameters:

\begin{equation}
FR=\frac{FR_{AA}(x_{AA})+FR_{SS}(x_{SS})+FR_{t1}(x_{t1})+FR_{t2}(x_{t2})+FR_{t3}(x_{t3})}{5},
\label{eq2}
\end{equation}

providing an estimation of the capability of hosting flares in each SG characterized by a particular configuration, size and fragmentation.

$FR$ is calculated for three different ranges of flare energy: C1.0 GOES class and greater (C1.0+), M1.0 class and greater (M1.0+) and X1.0 class and greater (X1.0+). 

Finally, following the Poisson statistics, the probability that a SGs host a flare is obtained by: 

\begin{equation}
p_{f}=1-exp(-FR).
\label{eq3}
\end{equation}

These probabilities are daily published at http://ssa.oact.inaf.it/oact/Flare\textunderscore forecasting.php, distinguishing for each SG visible on the solar disc and for each flare energy range (C1.0+, M1.0+ and X1.0+). We assume that the forecastings are valid for the subsequent 24 hours following the observation time, when approximately new observations and new measurements of the parameters are available. A screenshot of the web page containing the flare probabilities computed on September 6, 2017, is shown in Figure \ref{fig:res0}. As we can see, for each AR the NOAA number and the Catania Sunspot Group Number are indicted to identify the region where the flare may occur.

\subsection{An example of flare forecasting: AR NOAA 12673}

\begin{figure}
\centering
\includegraphics[scale=0.45]{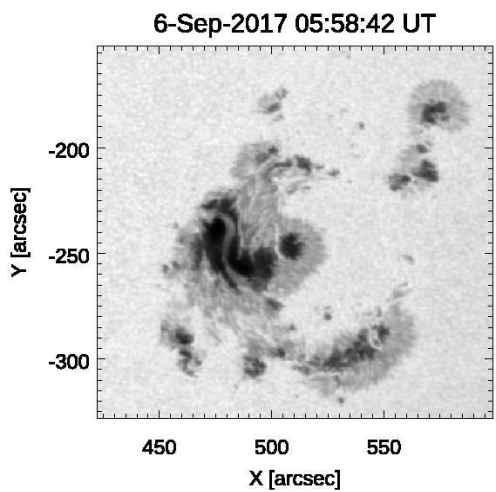}     
\caption{Continuum filtergrams taken by HMI/SDO at 617.3 nm. \label{fig:res1}}
\end{figure}

In order to illustrate the application of our method for a specific case, we consider the forecasting for the flare activity of the super active region (SAR) NOAA 12673 that appeared on the Sun during the first two weeks of September 2017 and unleashed more than 40 C-, about 20 M-, and 4 X-GOES class flares during its passage over the solar disc. In particular it hosted the strongest flare of the cycle 24 on September 6, 2017 \citep{Romano18, Romano19}. The SAR complexity is visible in Figure \ref{fig:res1}, where we show the continuum filtergram taken by HMI/SDO at 617.3 nm. 

The SAR NOAA 12673, as observed by INAF$-$OACt on September 6 at 6:50 UT, was formed by many pores and sunspots ($SS=45$), it was one of the largest SG of the solar cycle 24 ($AA=91$), its main sunspot was characterized by an asymmetric penumbra with a diameter greater then 2.5$^{o}$ ($t1=5$), the following spot was largest and the sunspot population density intermediate ($t2=5$), its configuration belong to the D class of the Zurich classification. On the base of these morphological characteristics we computed the $FR$ for each parameter and for each energy range, using Equation~\ref{eq1} (see Table \ref{table2}).

\begin{table}[h]
\caption{Flare rates for each parameter of the SAR NOAA 12673 and for their average.}
\label{table2}
\begin{tabular}{|c|c|c|c|}     
\hline                   
FR         & C1.0+ & M1.0+ & X1.0+ \\
\hline
$FR_{AA}(91)$    & 0.88 $\pm$ 0.04 & 0.49 $\pm$ 0.06 & 0.14 $\pm$ 0.04 \\
$FR_{SS}(45)$    & 0.79 $\pm$ 0.03 & 0.30 $\pm$ 0.03 & 0.06 $\pm$ 0.03 \\
$FR_{t1}(5)$     & 0.61 $\pm$ 0.02 & 0.22 $\pm$ 0.02 & 0.04 $\pm$ 0.01 \\
$FR_{t2}(5)$     & 0.49 $\pm$ 0.02 & 0.13 $\pm$ 0.01 & 0.02 $\pm$ 0.01 \\
$FR_{t3}(4)$     & 0.30 $\pm$ 0.01 & 0.05 $\pm$ 0.01 & 0.01             \\
$FR$             & 0.61 $\pm$ 0.02 & 0.24 $\pm$ 0.03 & 0.05 $\pm$ 0.02   \\
\hline
\end{tabular}
\end{table}

Then we computed their average by Equation~\ref{eq2} and the event probabilities for each range of flare energy by Equation~\ref{eq3}, obtaining the following results:

\begin{itemize}
\item 61\% for C1.0 class and greater (C1.0+)
\item 24\% for M1.0 class and greater (M1.0+)
\item 6\% for X1.0 class and greater (X1.0+)
\end{itemize}

These high percentages have been confirmed during the 24 hours subsequent to our forecast by 2 C-, 2 M- and 2 X-class flares, as reported in Table \ref{table3}.

\begin{table}
\caption{Flares occurred in the SAR NOAA 12673 during the 24 hours subsequent to the INAF$-$OACt flare forecasting.}
\label{table3}
\begin{tabular}{|c|c|c|c|c|}     
\hline                   
GOES class       & Date & Start Time (UT) \\
\hline
C2.7      & Sept 6 & 07:29  \\
X2.2      & Sept 6 & 08:57  \\
X9.3      & Sept 6 & 11:53  \\
M2.5      & Sept 6 & 15:51  \\
M1.4      & Sept 6 & 19:21  \\
C8.2      & Sept 7 & 06:19  \\

\hline
\end{tabular}
\end{table}

\section{Performance measures}

On the base of the flare records obtained by the GOES satellites and collected in the Space Weather Prediction Center Reports we evaluated the performance of our forecasting method. In Table~\ref{statistica} we summarize the statistics of the predictions considering data from 2002 until 2017.
We considered 8598 SGs. Among them the SGs hosting C1.0+, M1.0+ and X1.0+ class flares were 1841, 347 and 47, respectively. 
We constructed the probability distribution for forecasts, $f$, and observations, $o$. In Table~\ref{statistica} we report the properties of these distributions. $\left \langle p_{f} \right \rangle $ and $\left \langle o \right \rangle$ are the average of the forecast and observation probabilities over all observed SGs. We obtained the same values between them in the three energy ranges due to peculiarity of the method which is based effectively on the observed flare rates. However, the trend of $\left \langle p_{f} \right \rangle $ and $\left \langle o \right \rangle$ over the energy ranges depends on the smaller number of events of higher energy.

We also calculated the median for $p_{f}$ and the standard deviation for $p_{f}$ and $o$. With $\left \langle p_{f}|o=1 \right \rangle$ and $\left \langle p_{f}|o=0 \right \rangle $ we denote the average of the forecast probabilities over all observed SGs where at least one flare occurred and not, respectively. Over the same subsets the medians have been computed. Both the averages, with the corresponding standard deviations, and the medians show significant differences in the probabilities obtained for flaring and not flaring SGs.

The linear association describing the correlation between $p_{f}$ and $o$ provides already an indication of the reliability of our method for the three energy ranges (see \citet{Wheatland04} for more details).

\begin{table}
\caption{Verification statistics for the prediction of C+, M+ and X+ class flare events from data between 2002 and 2017.}
\label{statistica}
\begin{tabular}{|c|c|c|c|}     
\hline                   
Parameters                                          & C+ class flares &  M+ class flares  &  X+ class flares     \\
\hline
Sunspot groups                                      &       8598     &      8598          &      8598          \\
Sunspot groups with flares                          &       1841     &      347           &        47             \\
$\left \langle p_{f} \right \rangle              $  &      0.214      &     0.040         &       0.005           \\
$\left \langle o                  \right \rangle $  &      0.214      &     0.040         &       0.005           \\
Median $p_{f}                                    $  &      0.160      &     0.020         &       0.002           \\
$\sigma_{p_{f}}                                  $  &      0.143      &     0.049         &       0.010           \\
$   \sigma_{o}                                   $  &      0.410      &     0.197         &       0.074           \\
$\left \langle p_{f}|o=1 \right \rangle          $  &      0.337      &     0.127         &       0.036           \\
$\left \langle p_{f}|o=0 \right \rangle          $  &      0.136      &     0.037         &       0.005           \\
Median  $ p_{f}|o=1                              $  &      0.337      &     0.107         &       0.035           \\
Median  $ p_{f}|o=0                              $  &      0.136      &     0.018         &       0.002           \\
SD  $ p_{f}|o=1                                  $  &      0.170      &     0.085         &       0.023           \\
SD  $ p_{f}|o=0                                  $  &      0.112      &     0.043         &       0.009           \\  
Linear association                                  &      0.469      &     0.364         &       0.232           \\      
  \hline
\end{tabular}

\end{table}

To evaluate in detail our method, taking into account that it provides probabilistic forecasts, it has been necessary to determine a threshold probability to build a contingency table where any forecast probability over the threshold was considered to be a forecast for an event, and anything less was considered to be a forecast for non-event. The determination of the threshold, as we see in the following, depends on the adopted method of evaluation of the performance.

In fact, in a binary forecast method the contingency table is easily built: ``TP" is the number of true positives events, ``TN" is the number of true negative events, ``FP" is the number of false positive events and ``FN" is the number of false negative events. In that case the Rate Correct provides a first estimation of the forecast performance: $A_{for} =\frac{TP+TN}{N}$, where N is the total number of forecasts (N = TP + FP + FN + TN). For probabilistic forecasts, as in our case, TP, FP, FN, and TN depend on criteria to determine the threshold.

Usually, to avoid that the measure of accuracy can be misleading for very unbalanced event/no event ratio, it is necessary to normalize the performance to a reference forecast by using a so named skill score: 

\begin{equation}
Skill=\frac{A_{for}-A_{ref}}{A_{per}-A_{ref}}
\label{eq4}
\end{equation}

where $A_{per}$ and $A_{ref}$ are the accuracy of the perfect forecast and the reference method, respectively. We measured the accuracy of our probabilistic forecasts by two different skill scores (see \citet{Barnes16} for details): the Appleman’s Skill Score (ApSS) and the Hanssen \& Kuipers’ Discriminant (H\&KSS). 

According with the ApSS, the three accuracies used to calculate the skill score are the following:

\begin{equation}
A_{for}=\frac{TP+TN}{N},\\ A_{ref}=\frac{TN+FP}{N},\\ A_{per}=1
\label{eq5}
\end{equation}

According with the H\&KSS:

\begin{equation}
A_{for}=\frac{TP+TN}{N},\\ A_{ref}=\frac{TP}{TP+FN}-\frac{FP}{FP+TN},\\ A_{per}=1.
\label{eq6}
\end{equation}

\begin{figure}[h]
\centering
\includegraphics[scale=0.46]{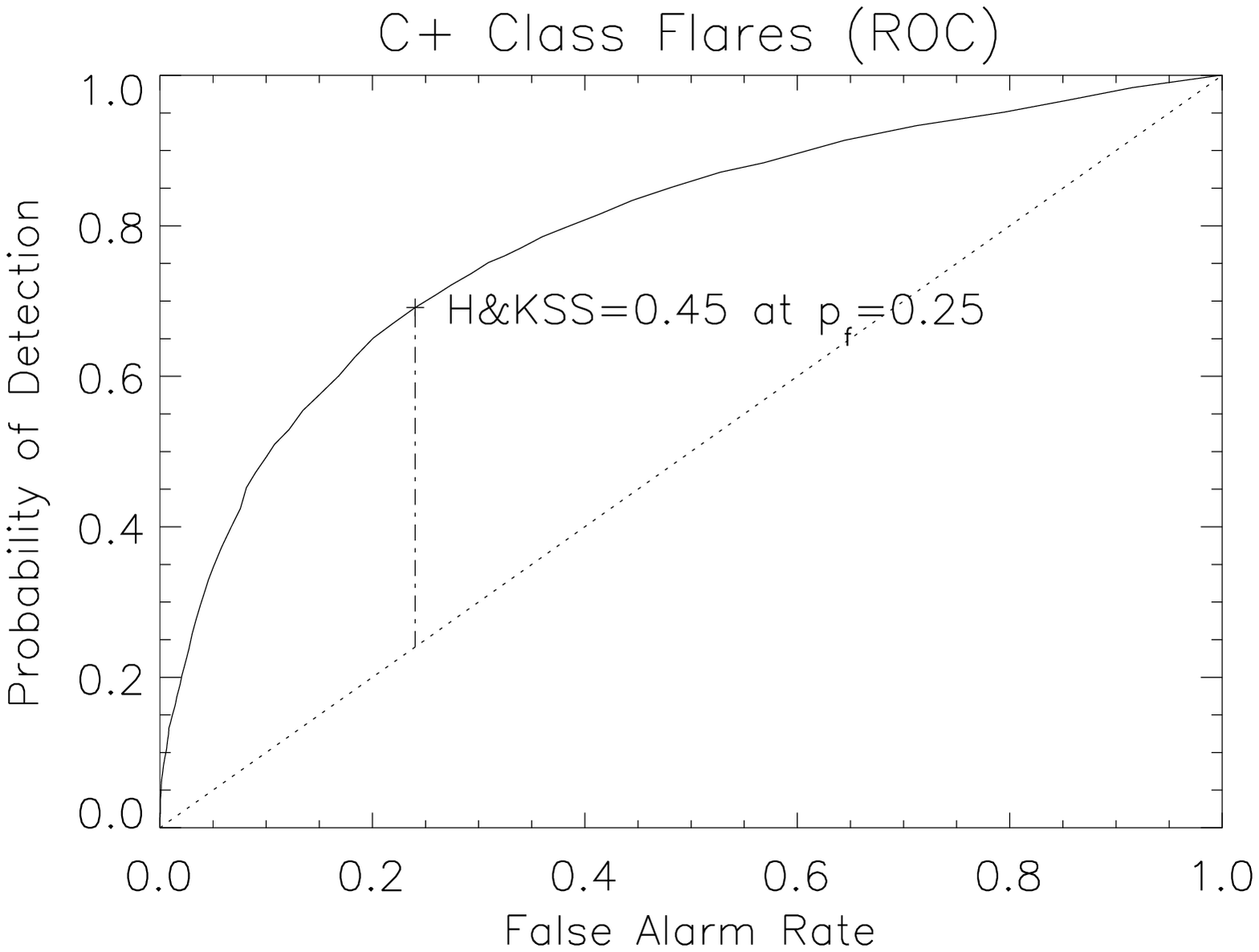}    
\includegraphics[scale=0.46]{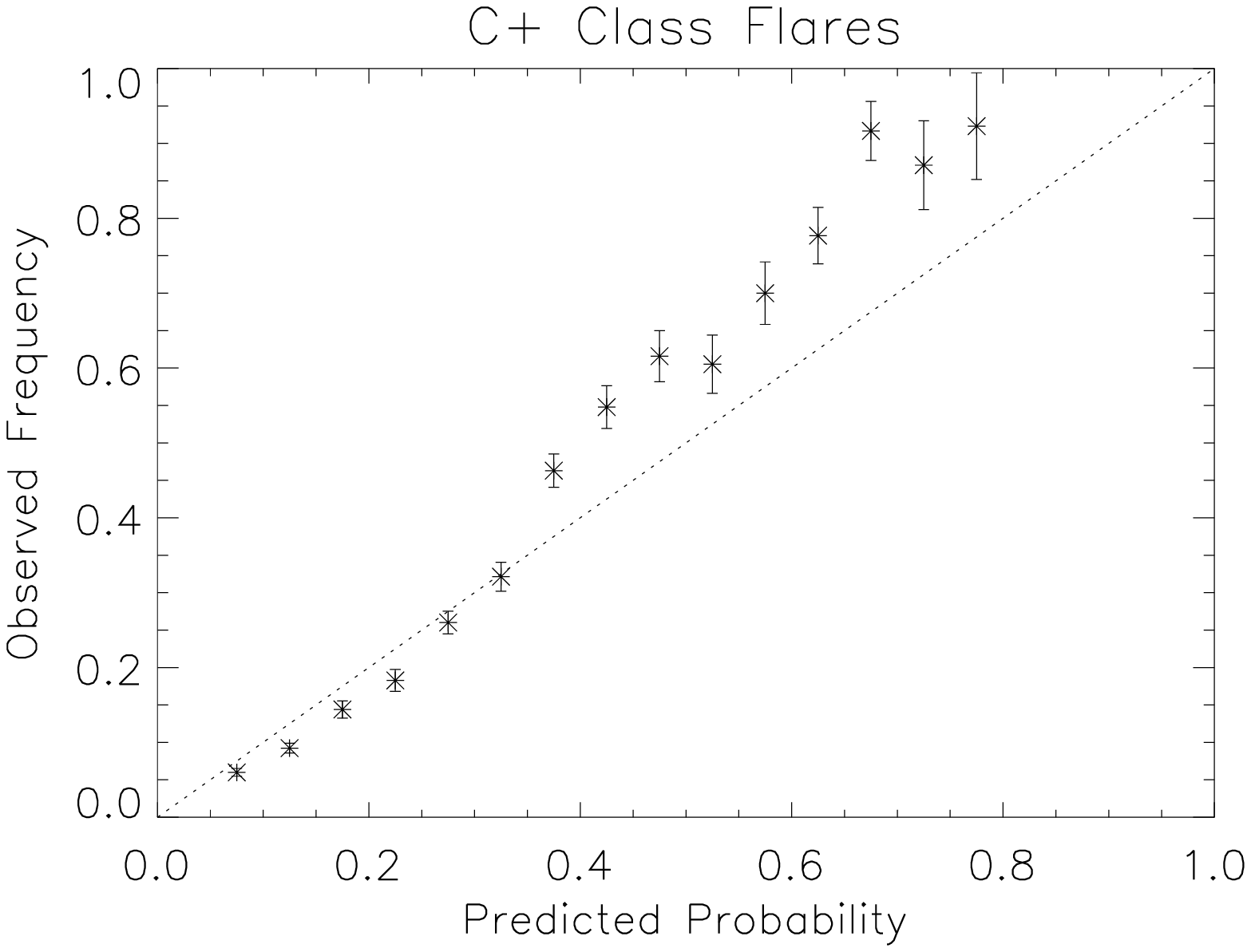}      \\
\includegraphics[scale=0.46]{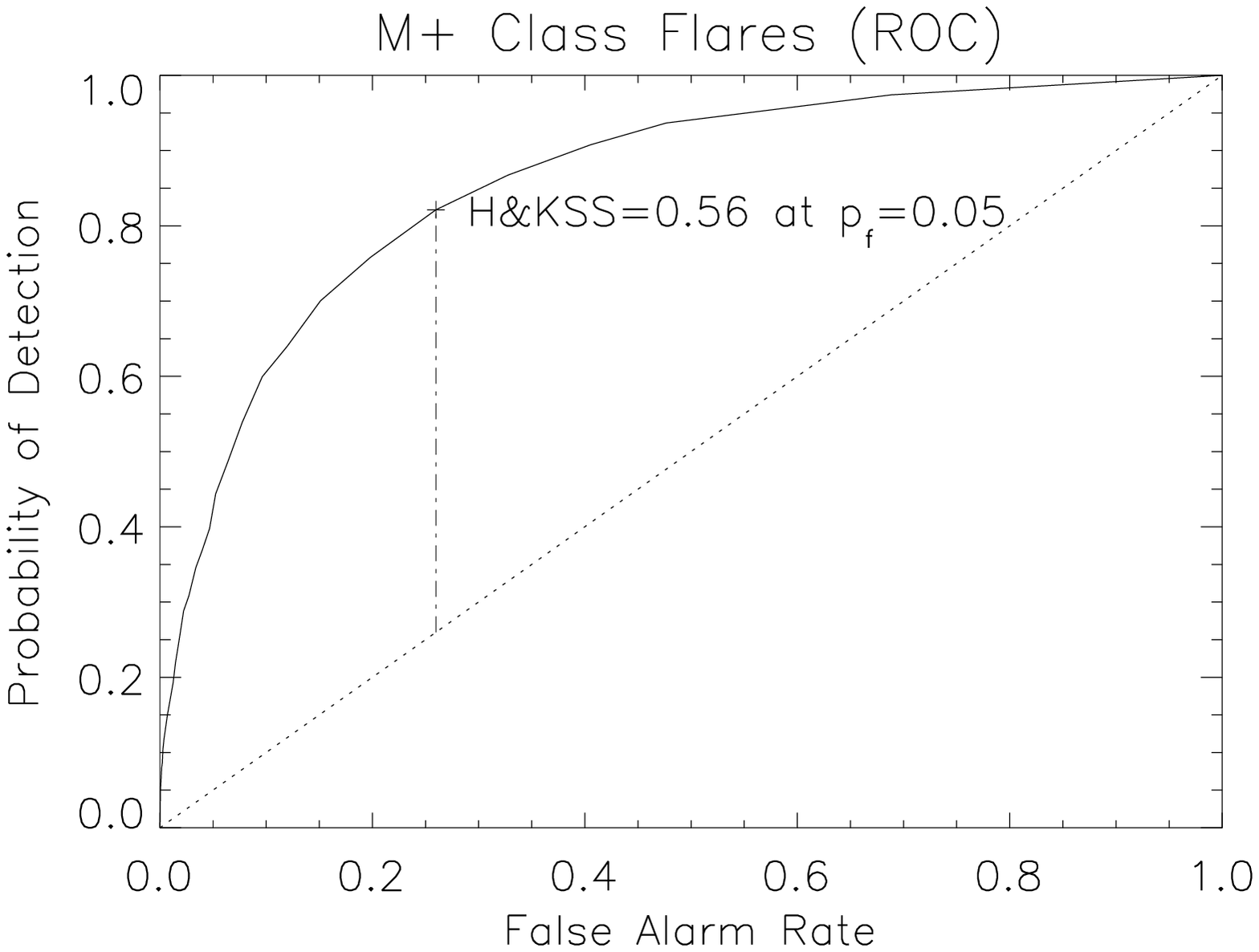}     
\includegraphics[scale=0.46]{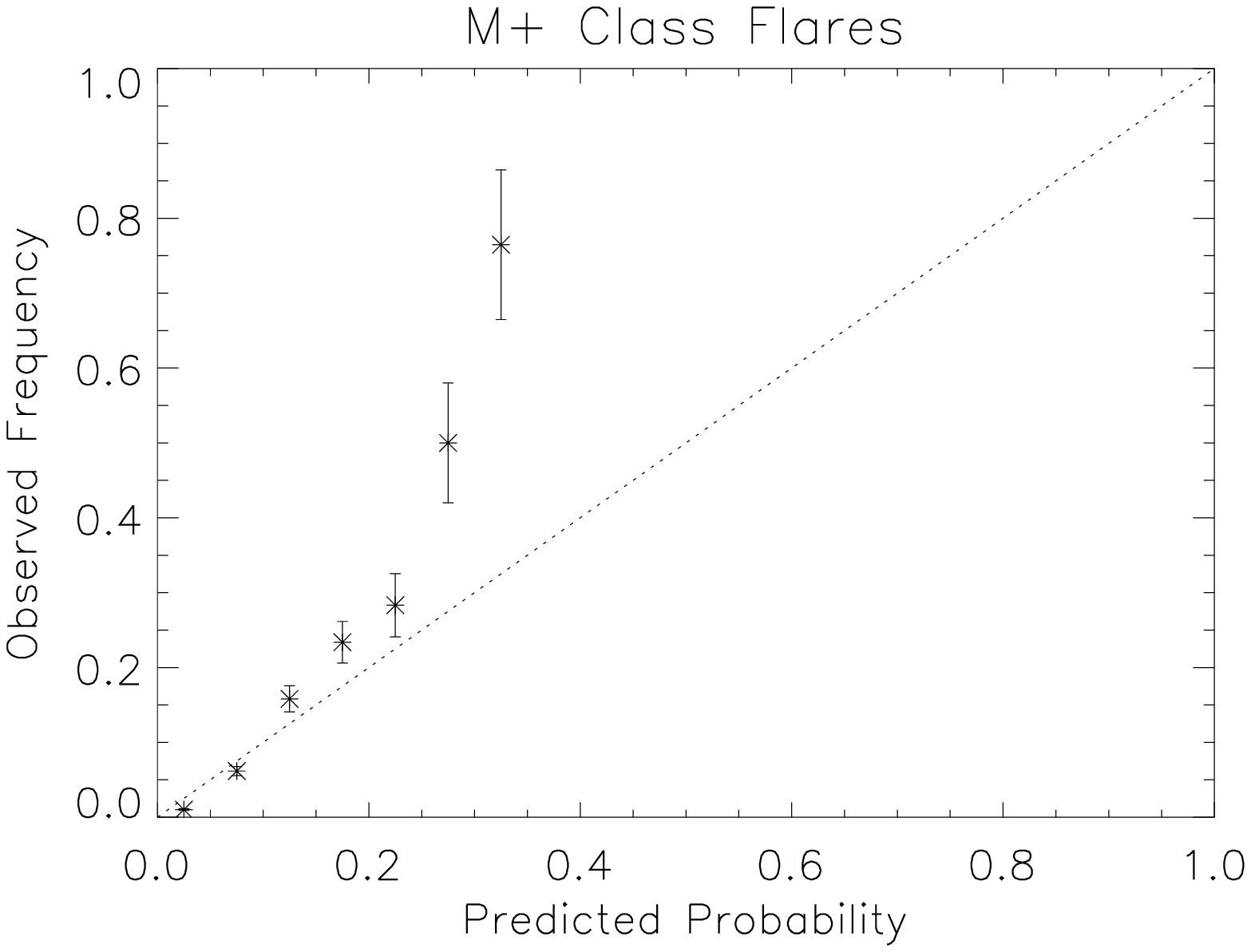}    \\
\includegraphics[scale=0.46]{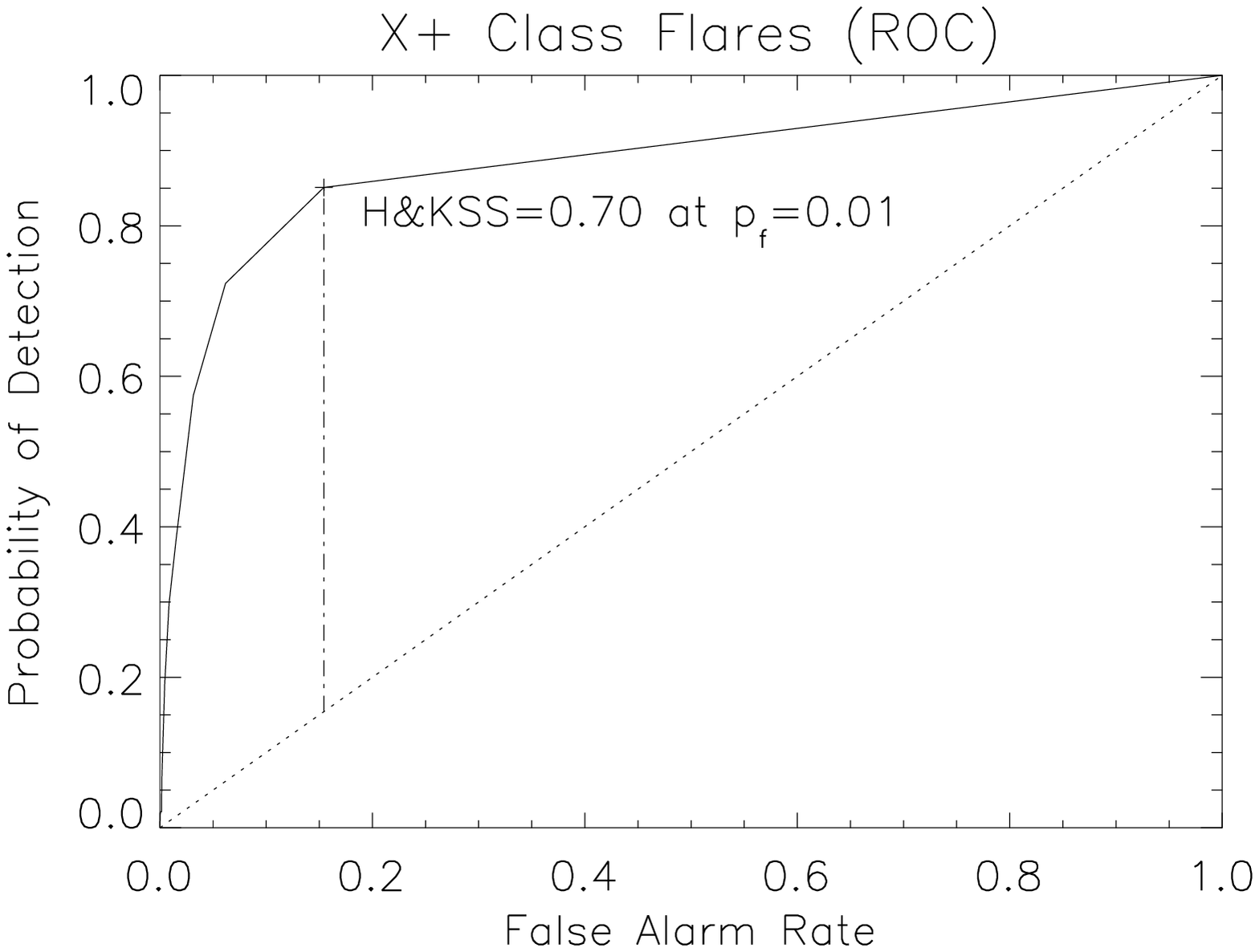}   
\includegraphics[scale=0.46]{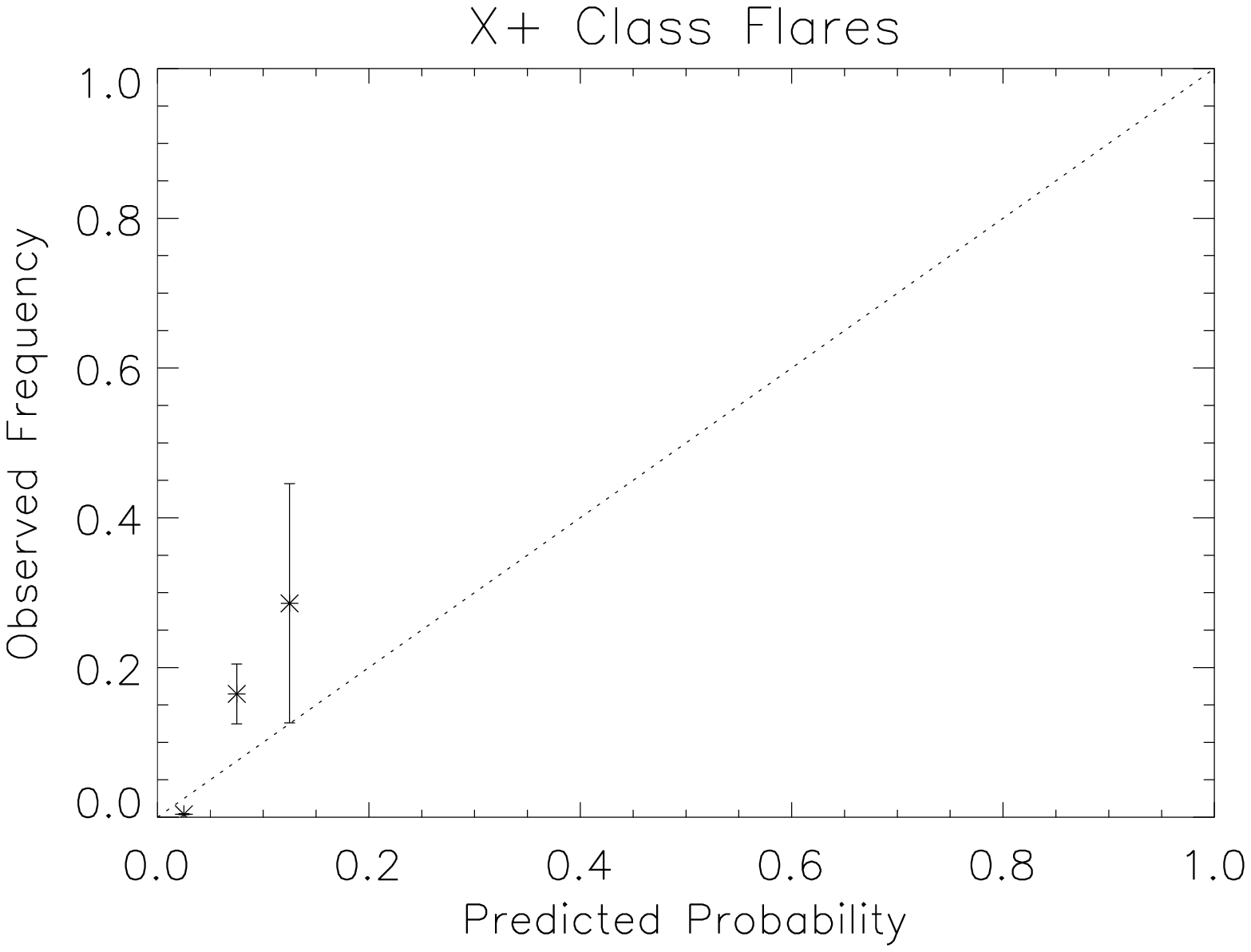}   

\caption{ROC curves (left panels) and reliability plot (right panels) for C1.0+, M1.0+ and X1.0+ class flares. $p_{f}$ indicates the threshold probability for generating the binary categorical classification which maximizes H\&KSS. \label{fig:res2}}
\end{figure}

We determined the threshold probabilities for each flare class and for each skill score and generated the binary categorical classifications, considering the values that maximize the two skill scores. These thresholds are reported in parenthesis in Table \ref{tab:metodi}. In particular for the H\&KSS we determined threshold probabilities from the so called Receiver Operating Characteristic ($ROC$) curves. The first term of $A_{ref}$ represents the probability of detection ($POD$), while the second term represents the false alarm rate ($FAR$). The $ROC$ curves show $POD$ as function of $FAR$ by varying the threshold above which a SG is predicted to produce a flare. We built a $ROC$ curve for each flare energy range and we determined the threshold probabilities which maximize the H\&KSS, considering the points of the curves that are more distant from the dotted diagonal line (see the left panels of Figure~\ref{fig:res2}). These thresholds are 0.25, 0.12 and 0.01 for C1.0+, M1.0+ and X1.0+, respectively.

Due to the fact that our flare forecasting method predicts the probability of a flare of a given class range rather than a binary, categorical forecast, a more appropriate measurement of accuracy is the mean square error (MSE):

\begin{equation}
A_{for}=MSE(p_{f},o)=\left \langle (p_{f}-o)^2 \right \rangle 
\label{eq7}
\end{equation}
where $p_{f}$ is the forecast probability, and $o$ is the observed outcome ($o=0$ for no event, $o=1$ for an event). In this case $A_{per} = 0$.

Using the climatological event rate as a reference forecast and its corresponding accuracy:

\begin{equation}
A_{ref}=MSE(\left \langle o \right \rangle ,o)=\left \langle (\left \langle o \right \rangle-o)^2 \right \rangle ,
\label{eq8}
\end{equation}

we computed the so called Barier Skill Score (BSS):

\begin{equation}
BSS=\frac{MSE(p_{f},o)-MSE(\left \langle o \right \rangle,o)}{0-MSE(\left \langle o \right \rangle,o)} .
\label{eq9}
\end{equation}

Therefore, we remark that $BSS$=0 corresponds to perfect performance.

In Figure~\ref{fig:res2} we report the reliability plots, which compare the predicted probabilities for all days sorted into bins of 0.05 width and the observed event rates. The associated uncertainties are calculated using the Bayesian estimation assuming binomial statistics (see Jaynes, 2003 for more details). It is noteworthy that the increasing size of the error bars with the increase of the predicted probabilities and the flare energy ranges reflects the decreasing sample size of the events. These plots allow us to understand the performance of the method for different flare probabilities. In fact, they show a slight tendency of over$-$prediction (point lying below x=y) for the smaller probabilities and a tendency of under$-$prediction (point lying over x=y) for the higher probabilities. We also note that there are few number of points in the reliability plot for X1.0+ flares due to the small number of event in the corresponding sample (see the fourth column of Table \ref{statistica}).  

All the results of the performances about our method are reported in Table~\ref{tab:metodi}. We see that all scores show a performance of our method better than the reference forecasts, although different skill scores emphasize different aspects of our performance. We note that $H\&KSS$ and $BSS$ are characterized by better values for higher flare class ranges (we remember that perfect performances should correspond to 1 and 0 for $H\&KSS$ and $BSS$, respectively). In particular, $H\&KSS$ varies from 0.45 for C1.0+ flares to 0.70 for X1.0+, while the threshold probabilities decreases from 0.25 to 0.01. This means that our method seems to get the best results for the more energetic events. However, we cannot neglect that $ApSS$ shows an opposite trend for the C1.0+ and M1.0+ classes. But, according to the results of \citet{Barnes16}, we see that the maximum forecast probability decreases with increasing event energy. This also explains the small values of $ApSS$ for larger events due to the small sample of events to be considered a predicted event in this categorical forecast.

\begin{table}[ht]
\caption{Performance results of ApSS, H\&KSS and BSS. In parenthesis we reported the threshold values which maximize the skill scores.}
 \label{tab:metodi}
 \centering 
  \begin{tabular}{cccc}
    \hline
    Flare class   & ApSS             &   H\&KSS          &    BSS         \\
    \hline
    C 1.0+        & 0.16 (0.40)      &  0.45 (0.25)      &    0.20        \\
    \hline
    M 1.0+        &  0.04 (0.27)     &  0.56 (0.05)      &    0.12         \\
    \hline
    X 1.0+        &  -               &  0.70 (0.01)      &    0.04          \\
   \hline
     \end{tabular}
\end{table}


\section{Discussion and conclusions} 
      
The flare forecasting method, applied to the daily INAF-OACt observations of the photosphere, provides an indication of the probabilities that each SG visible on the solar disc may host solar flares of C1.0+, M1.0+ and X1.0+ class. Taking into account the conclusions of \citet{Barnes16}, this method has the advantage to combine the characterization of each AR by some parameters and the application of a statistical analysis. The method requires only a daily observation of the solar disc at photospheric level, therefore, it is possible to collect all the information useful to compute the flare probabilities only exploiting a short temporal window of good weather. Moreover, the simple procedure to determine the parameter values from the photosphere observation and compute the flare rates allows to obtain the forecasts in few minutes after the observations. At the moment, the values of the parameters used as inputs for the method are obtained by eye. Once these values are determined, an automatized procedure consults the database and provides the flare probabilities for each AR visible on the solar disc. The probabilities are daily available at the web page of the INAF-OACT web site: http://ssa.oact.inaf.it/oact/Flare\textunderscore forecasting.php. 

The performance measures of our method show that we obtain the best results and more accurate forecasts for the stronger events. We do not achieve particularly high skill scores, suggesting that there is considerable room for improvement our method. The $BSS$, which is suitable for probabilistic forecasts, is equal to 0.04 for X1.0+ flares. This means that the photospheric configuration of a SG is already a valid indication of the possible occurrence of a strong flare, independently from its observations in corona. This is confirmed by $H\&KSS$=0.70 for X1.0+ flares with a threshold value of 0.01. These results support our method for its Space Weather purposes, in fact, the prediction of extreme events is particularly important to prevent 
serious damages to technological systems which pervade our life. Actually, $ApSS$ showed for C1.0+ and M1.0+ flares an opposite trend in comparison to the other skill scores, with a better performance for the sample containing also the weaker events. However, we impute this apparent conflicting results to the sensibility of $ApSS$ to the sample size \citep{Barnes16}. Also for this reason it was not possible to measure $ApSS$ for the X1.0+ flares. However, taking into account that our method determines the probabilities of a flare rather than a binary, categorical forecast, we are confident that the most accurate measures are those obtained by the $BSS$, which shows values closer to zero (perfect performance) as the energy of the predicted events increases.

The good results obtained by our method can be also confirmed by its comparison with other forecasting methods. One of the most promising method has been developed by \citet{Korsos15}. It is based on the weighted horizontal magnetic gradient ($WG_{M}$), defined between opposite polarity umbrae at the polarity inversion line of ARs. Studying the evolution of the unsigned magnetic flux, the distance between the area-weighted barycenters of opposite polarities and the $WG_{M}$, they are able to predict the flare onset time and assess whether a flare is followed by another event within about 18hr (see \citet{Korsos15} and \citet{Korsos19} for more details). We remark that their method is not based on a statistical base. Among the ARs used for the application of their method we can consider the AR NOAA 11504, which has been observed also by INAF-OACt. A M1.2 flare and a M1.9 flare occurred in this AR on June 13, 2012 at 13:17 UT and on June 14, 2012 at 14:35 UT, respectively. \citet{Korsos19} measured the relative gradient of the rising phase of the $WG_{M}$ and the relative gradient of the distance parameter of the converging motion for this AR. They obtained a $WG_{M}$ maximum value, $WG^{max}_{M}=0.55 x 10^{6} Wb/m$, which was followed by a less step decrease phase ended with the M1.2 flare ($WG^{flare}_{M}= 0.35 x 10^{6} Wb/m$) and the M1.9 flare ($WG^{flare}_{M}= 0.23 x 10^{6} Wb/m$). They also calculated the percentage differences ($WG^{\%}_{M}$) for the first and second flare which were $34\%$ and $61\%$, respectively. This values are in accordance with \citet{Korsos15}: in fact, if the $WG^{\%}_{M}$ is less then $42\%$ more flares are expected to follow, otherwise other flares are not expected. Actually, the AR NOAA 11504 showed another event after the first flare of M1.2 class and no other significant event after the second flare of M1.9 class.

The AR NOAA 11504 was observed by INAF-OACt on June 13, 2012 at 07:18. It was formed by many pores and sunspots ($SS=31$) with a large projected area $AA=54$. The main sunspot was characterized by a symmetric penumbra with a diameter grater then $2.5\degr$ ($t1=4$), the leading spot was largest and the sunspot population density intermediate ($t2=4$) and its configuration belong to the E class of the Zurich classification. On the base of the morphological characteristics the AR presented a probability of $13\%$ for the occurrence of M1.0+ flares. This percentage was widely higher than the threshold that maximizes the $H\&KSS$ for M1.0+ flares. The day after, on June 14, 2012 at 08:00 the morphological characteristics were the same: $t1=4$, $t2=4$ and $t3=5$, but the number of sunspots and pores ($SS=46$) and the projected area ($AA=72$) increased. The corresponding probability for M1.0+ flares was higher than before ($32\%$).

This comparison allows us to highlight some advantages of our method. For example, it is noteworthy that the percentages obtained by the observed parameters do not depend on the AR evolution and can be determined using a single observation of the photosphere. Although the method of \citet{Korsos15} is more sophisticated, it requires a continuity in the target observations that is not always guaranteed, especially by the ground based telescopes. Moreover, it seems that the higher probabilities of the INAF-OACt method correspond to higher intensity of the occurred flares. Therefore, we can conclude that, as for the $WG_{M}$ method, the INAF-OACt method could be able to estimate the likelihood of a subsequent flare of the same or larger energy.

However, we think that the limits of this method at different energy ranges may be ascribed to the indirect determination of the magnetic field configuration of the SGs. In fact, up today in our method the flare rates are determined on the base of the SGs characteristics apart from the polarities of pores and sunspots. Therefore, a possible improvement of the present version of the INAF-OACt method could be the implementation of the flare rates using also the McIntosh group classification and other useful magnetic field characterization parameters. 

A further margin of improvement of our forecasting method resides in the opportunity to weight the relative importance of each parameter. In fact, in Equation \ref{eq2} we compute the average among the flare rates. Instead, we think that the results obtained by a machine learning approach may provide a rating of each parameter and, consequently, may suggest a different multiplicative coefficient for each term of the average of Equation \ref{eq2} in order to get better performance.

It is noteworthy that we can improve our methods using the digital images acquired in the continuum at 656.78 nm and applying an automatic procedure to extract the considered parameters from those full disc images. In this way, we could be able to provide an upgrade of our forecasting in near real time according to the acquisition rate of the digital images (to date about 1 hour). Nevertheless, taking into account the usual dynamics and evolution of the ARs at photospheric level, we are confident that the main characteristics of the sunspot groups and, consequently, the corresponding flare rates do not change significantly in 24 hours.
   
Finally, we also plan to strength the statistics by the implementation of the database with data taken before January 2002. In fact, INAF-OACt has an archive of full disc drawings of the photosphere, containing the useful information for the SG characterization, since the beginning of the 20th century, and the GOES X-ray flux measurements at 1 - 8 Angstrom are available since 1986. We believe that the update of our databse will be a further opportunity to be more confident especially for the statistics of the strongest events, which are more interesting for their contribution to space weather conditions.


\begin{acknowledgements}

This project has received funding from the European Union's Horizon 2020 research and innovation programme under Grant Agreement No 824135 (SOLARNET).
The editor thanks two anonymous referees for their assistance in evaluating this paper.
\end{acknowledgements}



\end{document}